\documentclass[draft]{agujournal2019}
\usepackage{url} %this package should fix any errors with URLs in refs.
\usepackage{lineno}
\usepackage[inline]{trackchanges} %for better track changes. finalnew option will compile document with changes incorporated.
\usepackage{soul}

\draftfalse

\journalname{JGR: Space Physics}

\begin{document}

\newpage

\title{Alternating emission features in Io's footprint tail: Magnetohydrodynamical simulations of possible causes}

\authors{Stephan Schlegel \affil{1}, Joachim Saur \affil{1}}

\affiliation{1}{Institut für Geophysik und Meteorologie, Universität zu Köln, Cologne, Germany}

\correspondingauthor{Stephan Schlegel}{sschleg1@uni-koeln.de}

\begin{keypoints}
\item	Hall effect in Io's ionosphere produces Poynting flux morphology similar to observed alternating Alfv\'en spot street in Io footprint tail
\item	Alfv\'en wave travel time difference and asymmetries in Io's atmosphere are not sufficient to produce observed structures in Io footprint tail
\item	Io footprint tail emission inter-spot distance correlates with reflected Alfv\'en waves
\end{keypoints}

\begin{abstract}

Io's movement relative to the plasma in Jupiter's magnetosphere creates Alfv\'en waves propagating along the magnetic field lines which are partially reflected along their path. These waves are the root cause for auroral emission, which is subdivided into the Io Footprint (IFP), its tail and leading spot. 
New observations of the Juno spacecraft by \citeA{mura2018juno} have shown puzzling substructure of the footprint and its tail. In these observations, the symmetry between the poleward and equatorward part of the footprint tail is broken and the tail spots are alternatingly displaced.
We show that the location of these bright spots in the tail are consistent with Alfv\'en waves reflected at the boundary of the Io torus and Jupiter's ionosphere. Then, we investigate three different mechanisms to explain this phenomenon: 
(1) The Hall effect in Io's ionosphere, (2) travel time differences of Alfv\'en waves between Io's Jupiter facing and its opposing side and (3) asymmetries in Io's atmosphere. For that, we use magnetohydrodynamic simulations within an idealized geometry of the system. We use the Poynting flux near the Jovian ionosphere as a proxy for the morphology of the generated footprint and its tail.
We find that the Hall effect is the most important mechanism under consideration to break the symmetry causing the "Alternating Alfv\'en spot street". The travel time differences contributes to enhance this effect. We find no evidence that the inhomogeneities in Io's atmosphere contribute significantly to the location or shape of the tail spots.

\end{abstract}

\section{Introduction}
\begin{figure}
\begin{center}
\includegraphics[scale = 0.5]{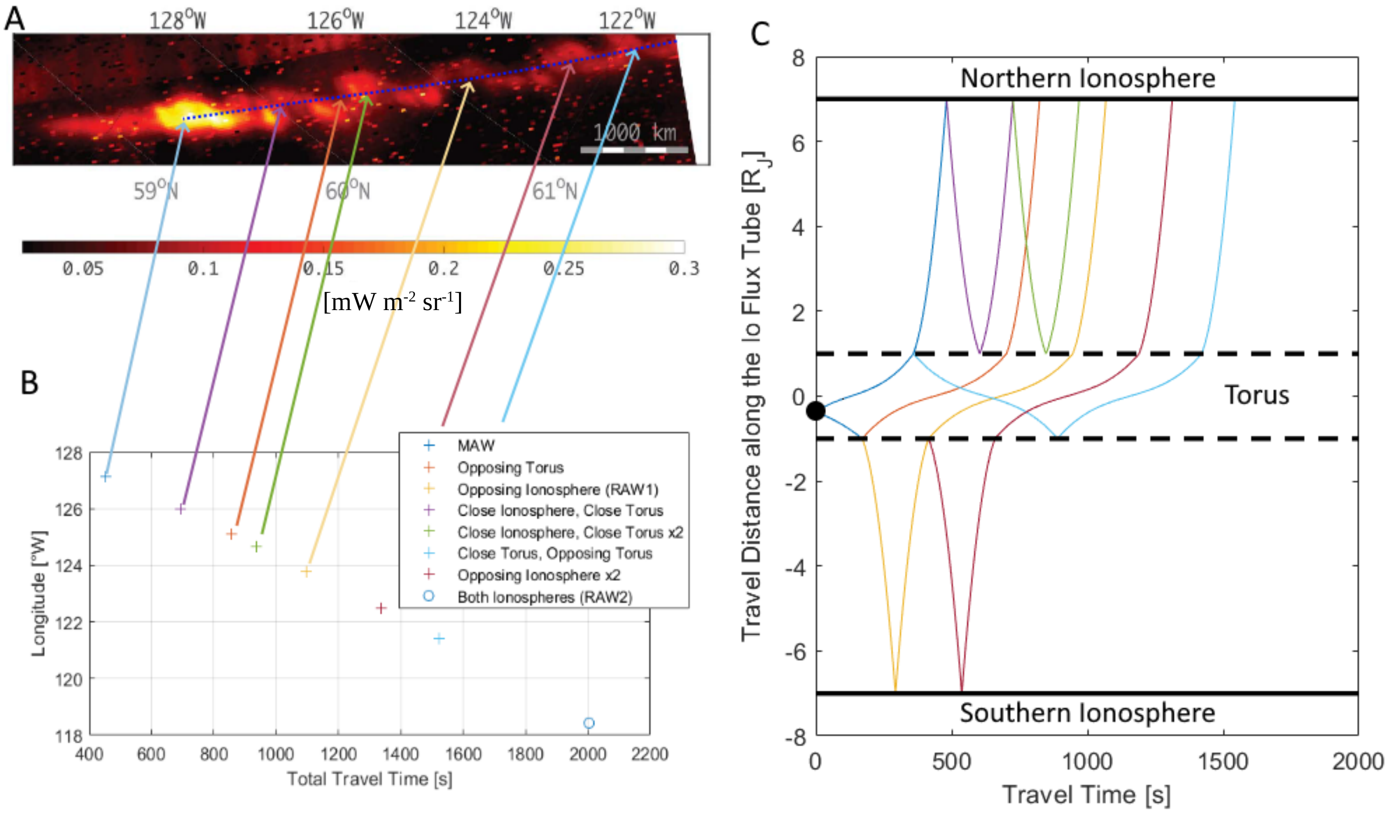}
\caption{(a) JIRAM infrared image taken of the IFP in the northern hemisphere (Taken from \citeA{mura2018juno}). Apart from the main spots at about 127°W, 59.5°N, multiple secondary spots are visible. These secondary spots are displaced perpendicular to the Io Footprint path and form an alternating Alfv\'en spot street. (b) Calculated total travel times of different Alfv\'en wave packages along the JRM09 curved magnetic field lines, depending on the reflection pattern and the western longitude of their corresponding footprint. The arrows map the calculated position of the footprints on the Juno infrared image. (c) Illustration of the reflection patterns of the footprints in (b). The color of the lines in (c) corresponds to the color of the markers in (b). The position of Io is shown as a small circle and the position of maximum reflection at the torus boundary is shown as dashed line.}
\label{fig:Mura}
\end{center}
\end{figure}

Jupiter's strong aurora is generated by particles precipitating onto the planet's ionosphere. Apart from polar emission, the main auroral oval, and diffuse emissions equatorward (e.g. \citeA{grodent2018jupiter}), very distinct features are the footprint spots and tail emissions that are associated with the Galilean moons. Of those footprints, Io generates the brightest one in ultraviolet \cite{clarke1996far}, visible light \cite{vasavada1999jupiter}  and infrared \cite{connerney1993images,connerney2000h3+,mura2018juno}. The footprints of Europa and Ganymede are  fainter, however well detectable. The Callisto footprint, which is very close to the auroral main emission, has only been  likely  detected in two occasions \cite{bhattacharyya2018evidence}.
Observations of the shape, intensity and position of these footprints including their leading and tail emissions can be used as a diagnostic to better understand the moon-planet interaction.\\
After detection of the dense Io torus, Alfv\'en wing models have been developed in the magnetohydrodynamic (MHD) framework \cite{neubauer1980nonlinear,goertz1980io} to explain this interaction and the location of the related footprint emission, the secondary spots in the tail and the leading spot \cite{bonfond2008uv}. Withing these Alfv\'en wing models,  Alfv\'en waves are generated when the corotating plasma in Jupiter's inner magnetosphere exchanges momentum with Io's atmosphere. These waves travel along their characteristics towards Jupiter, where they accelerate particles above Jupiter's ionosphere \cite{crary1997generation, damiano2019kinetic, szalay2018situ, szalay2020new}.  The accelerated particles travel along the field lines in both directions, creating auroral footprints on both hemispheres \cite{hess2010power}. Depending on Io's position inside the torus, particles generated at one hemisphere of Jupiter can travel towards the other faster than the main Alfv\'en wing (MAW). These transhemispheric electron beams (TEB) generate leading spot emissions \cite{bonfond2008uv}. The Alfv\'en waves are also reflected at the torus boundary and ionosphere of Jupiter \cite{neubauer1980nonlinear,bagenal1983alfven, gurnett1983ion} and create tail spots and continuous tail emission. Magnetohydrodynamical simulations have been carried out by \citeA{jacobsen2010location} to explain the transition from tail spots to continuous spot emission for Io's varying position in the Io plasma torus. They showed that the transition is controlled by the effect of high interaction strength and non-linear reflection. Another explanation of the tail emission was suggested to be due to the acceleration of the plasma in Io's wake to corotational speeds  by $\mathbf{j} \times \mathbf{B}$ forces \cite{hill2002jovian, delamere2003momentum}. The related current in Io's wake connects via field-aligned currents to Pedersen currents in Jupiter's ionosphere. The field aligned currents are argued to lead to quasi-static electric potentials that accelerate electrons towards Jupiter. However, the arising mono-energetic electron distributions are not consistent with the ultraviolet observations of the IFPT \cite{bonfond2009io} and we assume the mass-loading contributing a relatively small amount to  the total momentum exchange in the system as discussed by \citeA{bonfond2017tails}.\\
With the Juno spacecraft in orbit around Jupiter, the Jovian InfraRed Auroral Mapper (JIRAM) has made high resolution infrared images of the footprints and tails of Io \cite{mura2018juno, moirano2021morphology} as well as Europa and Ganymede \cite{moirano2021morphology}, which includes features that can not be explained by current models.  One of the observations of the Io footprint and its tail spots is shown in Figure \ref{fig:Mura}A.  The tail shows an alternating substructure, indicated by the arrows. In these images, the tail spots are alternatingly displaced polewards and equatorwards from their track, calculated using the JRM09 magnetic field model \cite{connerney2018new}. We refer to these features as "Alternating Alfv\'en Spot Street" (AASS). Additionally, a bifurcation of the tail has been observed in the infrared observations \cite{mura2018juno} as well as in electron flux signatures \cite{szalay2018situ}. Later observations also revealed substructures that are fixed within Jupiter's rest frame, for which \citeA{moirano2021morphology}suggested a feedback mechanism between Jupiter's ionosphere and the magnetosphere. The aim of our work is not a detailed reproduction of all features of Io's footprint and associated tail features including the substructures reported in \citeA{moirano2021morphology} or the bifurctions in \cite{mura2018juno}, but a basic investigation of the cause of the alernating structures, which are fixed in Io's rest frame. \\
 In this work, we first evaluate whether the downstream distances of the spots in the AASS are consistent with reflected and refracted Alfv\'en waves at the Io torus boundary and Jupiter's ionosphere (Figure \ref{fig:Mura} B and C). Afterwards, we use Hall-MHD simulations to study the evolution of Alfv\'en waves through the Jovian inner magnetosphere generated at Io and the morphology of the emissions they would produce. Here, we investigate three different mechanisms that could break the poleward/equatorward symmetry of the footprint and its tail and create alternating structures similar to the observed pattern downstream the main Alfv\'en wing spot. 
 The first mechanism is the Hall effect in Io's atmosphere simulated for different ratios between the atmospheric Pedersen and Hall conductances. As a second mechanism, we consider the role of the Alfv\'en wave travel time between Io's Jupiter facing side and its opposing side. For the final mechanism, we implement asymmetries in Io's atmosphere. The simulation results including each of these mechanisms will be described and discussed in sections \ref{sec:Hall}, \ref{sec:TravelTime} and \ref{sec:Atmosphere}, respectively.

\section{Model and Methodology}\label{sec:Model}

 To understand the cause of the alternating Alfv\'en spot street, we perform numerical simulations with a single fluid Hall-MHD model to reproduce the Alfv\'en wing and its reflection and refraction pattern. For the used plasma temperature of 50~eV, the gyroradii of the two most abundant ion species near Io, O$^+$ and S$^+$ is about 2~km with a cyclotron frequency of 10~s$^{-1}$ and 5~s$^{-1}$, respectively. Therefore, the period of the ion cyclotron motion is much smaller than the convection time scale of $\tau = 64$~s for the plasma to bypass Io. Since the characteristic length scale of $R_{Io} = 1822$~km is much larger than the gyroradii, we assume the MHD approach to be applicable in Io's vicinity. In the high latitudes, the Alfv\'en velocity approaches to the speed of light and displacement currents take effect.
However, we adapted the geometry of the model domain as well as the density and magnetic field strength so that the Alfv\'en velocity is kept well below the speed of light and Alfv\'en wave travel times are consistent with those calculated from density and magnetic field models \cite{hinton2019alfven}.

\subsection{The Reference Model} 
\begin{figure}
\begin{center}
\includegraphics[scale = 0.7]{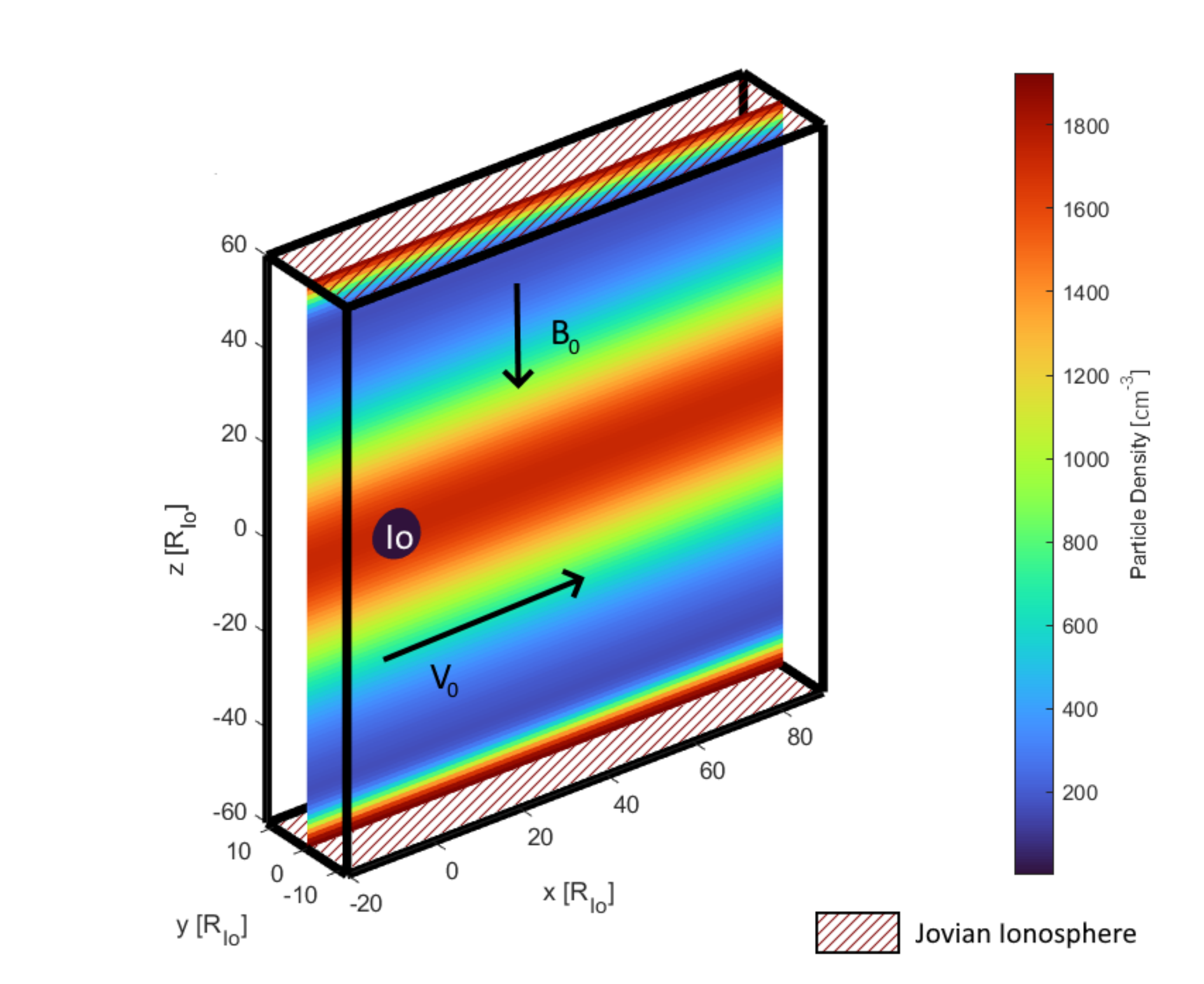}
\caption{Sketch of the model within which propagation of Io's Alfv\'en wings are calulcated. The magnetic field is assumed to be homogeneous and straightened and  points at negative z direction, while the incoming plasma flow is perpendicular to it in positive x direction with the relative velocity $v_0$ between Io and its surrounding plasma. The density decreases with distance z from the torus center and increases close to the  northern ($z$~=~+60~$R_{Io}$) and southern ($z$~=~-60~$R_{Io}$) Jovian ionosphere. Values of the reference model are given in table \ref{tab:properties}.} \label{fig:model}
\end{center}
\end{figure}

To model the generation of the Alfv\'en wings with the occurring reflections and refraction at the torus boundary as well as the Jovian ionosphere, the full extend of the flux tube needs to be covered by the simulation domain. To simplify the geometry we straightened the system similar to \citeA{jacobsen2007io,jacobsen2010location} as shown in Figure \ref{fig:model}. The x axis of the used Cartesian coordinate system is parallel to the incoming plasma flow and the z axis indicates the distance from the torus center anti-parallel to the background magnetic field. The y-axis completes the right hand coordinate system and points towards Jupiter with the center of Io at (0,0,0). The simulation is carried out in the reference frame of Io, therefore Io is stationary. The inflow plasma velocity is set to the velocity relative to Io  $v_0 = 57$~km/s. The homogeneous and constant background magnetic field $B_0 = 1720$~nT along the z-direction corresponds to the value of the Jovian magnetic field strength in the vicinity of Io.

\subsubsection{The density model} \label{sec:density}

To include realistic reflections and refraction of the Alfv\'en waves in the inner Jovian magnetosphere, the model needs to match the Alfv\'en velocities and travel time along the flux tube, because refraction angle and reflection strength are dependent on the ratio of the Alfv\'en velocities \cite{wright1987interaction}.  Since the magnetic field strength is constant in the model domain, the density gradient parallel to the magnetic field lines needs to be adapted accordingly. The density model,   given by 
\begin{equation}
    \rho(z) = \rho_0 + \rho_T(z) + \rho_I(z) \label{eq:density}
\end{equation}
 consists of a part corresponding to the torus $\rho_T$, an increase of density near the Jovian ionospheres $\rho_I$, and a minimum density in the inner magnetosphere $\rho_0$ to avoid relativistic Alfv\'en velocities that are not covered by our model. For the torus, we use an exponentially decreasing density
\begin{equation}
    \rho_T(z) = m~n_T~\textrm{exp}\left[ -\frac{z^2}{H_T^2} \right],
\end{equation}
with a central torus number density of $n_T = 1.8 \cdot 10^9$~m$^{-3}$ and a scale height of $H_T = 25 R_{Io}$  in agreement with the findings of the density and scale height of the ribbon and warm torus by \citeA{phipps2018distribution} and \citeA{dougherty2017survey}.  The used particle mass of $m = 24$~amu is calculated as approximated average ion mass of the system (e.g. Dougherty et al. 2017). Near the Jovian ionosphere we added a second density term $\rho_I$ to implement reflections at the northern and southern ionosphere. 
\begin{equation}
    \rho_I(z) = m~n_I~\textrm{exp}\left[ -\frac{z_I - |z|}{H_I} \right].
\end{equation}
 The chosen ionospheric scale height of $H_I = 0.4 R_{Io}$ \cite{su2006io} ensures a high reflection coefficient close to 1. The ionospheric maximum density of $n_I = 10^{11}$~m$^{-3}$ guarantees that most of the Alfv\'en wave is reflected before it leaves the model domain at $z_I = 60 R_{Io}$. 
For the inner magnetospheric minimum density, we chose $\rho_0 = m \cdot 10^8$~m$^{-3}$. This limits the maximum Alfv\'en velocity to 769 km/s, which neglects effects of the low density region, that has been observed by Juno Waves instrument \cite{elliott2021high, sulaiman2021inferring} and JADE (e.g. \citeA{allegrini2021electron}) or predicted by models (e.g. \citeA{su2006io}). However, this region has essentially no effect on the overall travel time of the Alfv\'en waves, since the Alfv\'en velocity approaches the speed of light. Given our simplified magnetic field model for the 3D MHD simulation part of this work, we chose a density model which on the one hand ensures strong reflections at the ionospheric boundary as well as gradual reflections at the torus boundary. On the other hand the model fits the estimated Alfv\'en wave travel time from the torus center to the Jovian ionosphere of $t_0 = 365$~s  which is according to the values calculated by \citeA{bagenal1983alfven} and \citeA{hinton2019alfven} for a one way Alfv\'en wave trip. 

\subsubsection{Parameterization of Io}
Io itself is implemented as a neutral gas cloud with constant neutral gas density $n_n(r<R_{Io}) = n_0$ inside and exponential decreasing density $n_n(r > R_{Io}) = n_0 \textrm{exp}(-r/H)$ outside with a constant scale height $H = 180$~km. The gas cloud acts as a conductive obstacle in the plasma flow, generating Alfv\'en waves. The Pedersen conductivity $\sigma_P$ and the Hall conductivity $\sigma_H$ for the gas cloud can be calculated as
\begin{equation}\label{eq:pedersen_con}
    \sigma_P = \frac{n e}{B}\frac{\nu \omega}{\nu^2 + \omega^2} 
\end{equation}
and 
\begin{equation}\label{eq:hall_con}
    \sigma_H = \frac{n e}{B}\frac{\nu^2}{\nu^2 + \omega^2},
\end{equation}
while $\omega \approx10$~s$^{-1}$ is the ion cyclotron frequency.  Here, only the collisions of ions with the neutral particles are considered. The interaction can be characterized using the Pedersen and Hall conductances \cite{saur1999three} given by
\begin{equation}
    \Sigma_{P,H} = \int \sigma_{P,H} ds,
\end{equation}
with the integration performed along the magnetic field line. Values for the Io-centered, i.e. $x=y=0$, Pedersen and Hall conductances are shown in table \ref{tab:properties}. 

\begin{table}
\caption{Properties of the reference simulation}\label{tab:properties}
\centering
\begin{tabular}{l c c}
\hline
Property & Symbol & Value  \\
\hline
Io Radius & $R_{Io}$ & 1822 km \\
Background Magnetic Field & $B_0$ & 1720~nT   \\
Inflow Plasma Bulk Velocity & $v_0$ & 57~km~s$^{-1}$ \\
Convection Time & $\tau$ & $2 R_{Io} / v_0 = 64$~s\\
Alfv\'en Travel Time & $t_0$ & 365~s \\
Central Torus Plasma Number Density & $n_T$ & $1.8\cdot10^9$~m$^{-3}$ \\
Central Torus Alfv\'en Velocity & $v_{A,0}$ & 181 km~s$^{-1}$\\
Central Torus Alfv\'en Mach Number & $M_{A}$ & 0.31\\
Background Pressure & $p_0$ & 29~nPa\\
Ion Cyclotron Frequency$^a$ & $\omega$ & 10~s$^{-1}$ \\
Io Neutral Gas Scale Height & $H$ & 200~km \\
Central Io Neutral Gas Density & $n_0$ & $3.3\cdot10^{12}$~m$^{-3}$ \\
Central Ion-Neutral Collision Frequency & $\nu_0$ & 1.14 s$^{-1}$\\
Io's Pedersen Conductance & $\Sigma_P$ & 50~S\\
Central Torus Alfv\'en Conductance & $\Sigma_A$ & 4.3~S\\
Plasma Beta & $\beta$ & 0.01 \\
\hline
\multicolumn{2}{l}{$^{a}$ For an $O^+$ ion}
\end{tabular}
\end{table}

\subsection{MHD Model Equations}  \label{sec:MHDModelEquations}
For the simulation we used the PLUTO code \cite{mignone2007pluto} to solve the system of ideal MHD equations (\ref{eq:continuity}) - (\ref{eq:energy}) with added collision terms (see also \cite{schunk1975transport, chane2013modeling, blocker2018mhd}) in the momentum and energy equation as well as the Hall term in the induction equation.

\begin{linenomath*}
\begin{equation}\label{eq:continuity}
\frac{\partial \rho}{\partial t} + \nabla \cdot (\rho \mathbf{v}) = 0
\end{equation}
\begin{equation}
\rho \frac{\partial \mathbf{v}}{\partial t} + \rho \mathbf{v} \cdot \nabla \mathbf{v} = - \nabla p + \frac{1}{\mu_0}(\nabla \times \mathbf{B}) \times \mathbf{B} - \nu \rho \mathbf{v}
\end{equation}
\begin{equation}
\frac{\partial \mathbf{B}}{\partial t}  = \nabla \times \left(\left(\mathbf{v} - \frac{m}{\mu_0 e} \frac{\nabla \times \mathbf{B}}{\rho} \right) \times \mathbf{B} \right)\label{eq:induction}
\end{equation}
\begin{equation}\label{eq:energy}
\frac{\partial \epsilon}{\partial t}  = - \nabla \cdot (\epsilon\mathbf{v}) - p \nabla \cdot \mathbf{v} + \nu\left(\frac{1}{2}\rho v^2 - \epsilon\right)
\end{equation}
\end{linenomath*}
with the mass density $\rho$, the particle mass $m$, the bulk velocity $\mathbf{v}$, the magnetic field $\mathbf{B}$, the elementary charge $e$ and the specific internal energy $\epsilon$, which is related to the pressure $p$  by 
\begin{equation}
    \epsilon = \frac{3}{2} p.
\end{equation}
The collision frequency can be calculated as $\nu(\mathbf{r}) = \sigma_C n_n(\mathbf{r}) v_{rel}$ after \citeA{saur1999three} with the collisional cross section $\sigma_C=2\cdot10^{-20}\textrm{m}^{2}$ and the neutral gas density $n_n$. 
Since the simulations are carried out in the rest frame of Io, the relative neutral gas velocity is zero and the relative velocity $v_{rel}$ between plasma and neutral gas simplifies to the plasma bulk velocity $v_{rel} = |\mathbf{v}|$.

\subsubsection{Boundary Conditions and Numerical Grid}
 The whole domain extends from -20 to 100 $R_{Io}$ in x direction, making sure that multiple reflection take place inside the model domain, -12 to 12 $R_{Io}$ in y direction and -65 to 65 $R_{Io}$ in z direction, with Io at  $\mathbf{r} = (0,0,0)$. The total simulation time was set to 80~min to ensure that the perturbations travel through the whole domain, while reducing possible reflections at the x = 100 $R_{Io}$ boundary.
The grid spacing is set constant in all directions to  $\Delta x = \Delta y = \Delta z = 0.2 R_{Io}$, leading to a total amount of [601 x 121 x 651] grid cells. 
The inflow velocity is kept constant and set to $\mathbf{v}(x_0) = v_0 \mathbf{e}_x$ at the inflow boundary $x_0 = -20 R_{Io}$. At all other boundaries, an outflow boundary condition is applied. This means that the gradients through the boundary are set to zero for magnetic field, density, velocity and pressure. For the initial conditions, the background velocity in the whole domain is set to $\mathbf{v}_{ini} = v_0 \mathbf{e}_x$, while the pressure is set constant to $p_0$. The values of the initial conditions are shown in table \ref{tab:properties}.

\subsection{Reference Simulation}

\begin{figure}
\begin{center}
\includegraphics[trim = 0cm 4cm 0cm 4cm, clip = true ,scale = 0.7]{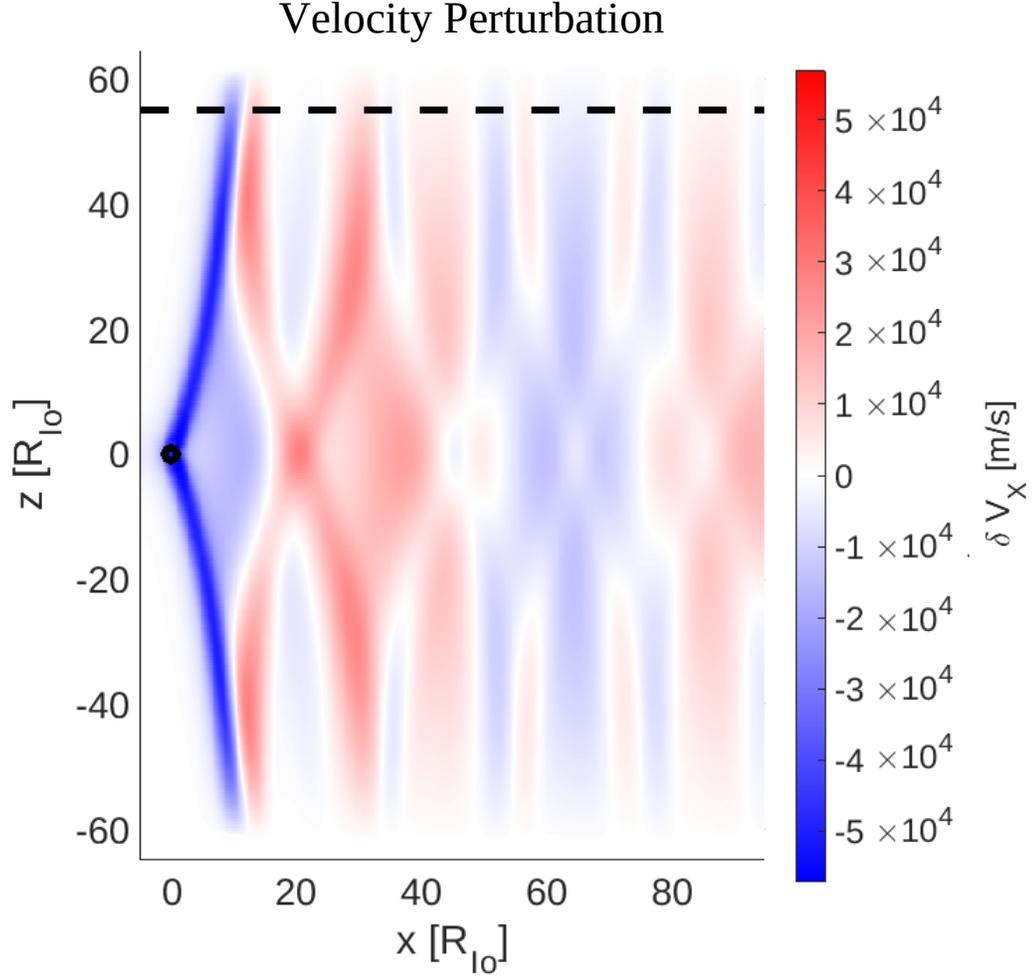}
\caption{Velocity field parallel to the incoming plasma flow of the reference simulation in the x,z plane (y~=~0). The colorbar is adjusted to show unperturbed plasma with a velocity of $v = v_0 = 57$~km~s$^{-1}$ in white. Red colors show a positive velocity perturbation (accelerated plasma) while blue show a negative velocity perturbation (decelerated plasma). The analysis plane of the northern Jovian ionosphere is shown as dashed line.}
\label{fig:reference}
\end{center}
\end{figure}

To evaluate the different mechanisms which could break the symmetry in the Io Footprint tail (IFPT), we compare the simulations (sections \ref{sec:Hall}, \ref{sec:TravelTime} and \ref{sec:Atmosphere}) with a reference simulation that does not include any of those mechanisms and therefore does not show any asymmetries.  The result of the reference simulation is shown in Figure \ref{fig:reference}. The main Alfv\'en wing (MAW) is shown as a strong deceleration of the plasma starting at Io (x~=~z~=~0). The interaction can be characterized using the interaction strength $\bar{\alpha}$, which is the decrease of  the background electric field $E_0$ towards the electric field inside the Alfv\'en wing $E_{AW} = -\mathbf{v} \times \mathbf{B}$  \cite{saur2013magnetic}. It can be approximated using the ratio between Pedersen conductance and Alfv\'en Conductance $\Sigma_A$ \cite{southwood1980io, saur1998interaction}.
\begin{equation}
    \bar{\alpha} = 1 - \frac{E_{AW}}{E_0} \approx \frac{\Sigma_P}{\Sigma_P + 2 \Sigma_A}.\label{eq:interaction_strength}
\end{equation}
The Alfv\'en conductance is calculated after \citeA{neubauer1980nonlinear} as
\begin{equation}
    \Sigma_A = (\mu_0 v_A (1 + M_A^2))^{-1}.
\end{equation}
This corresponds to a decrease of electric field of about $85\%$ inside the Alfv\'en wing in our model, which can also be seen in the velocity decrease in Figure \ref{fig:reference}. The velocity perturbation changes sign at a negative Alfv\'en velocity gradient, i.e. at the Jovian ionosphere, while it keeps sign at a positive Alfv\'en phase velocity gradient, i.e. at the transition between torus and inner Jovian magnetosphere at $z \approx 25 R_{Io}$. For the magnetic field perturbation, the opposite is the case, meaning a sign reversal at a positive Alfv\'en phase velocity gradient. Therefore, at every reflection the direction of Poynting flux is reversed \cite{wright1987interaction}. Since the boundary of the torus is a rather smooth gradient, the reflection is blurred. Together with the rather hard reflection at the Jovian ionosphere, a complex reflection pattern develops.  The overall reflection pattern and velocity distribution is comparable to the findings of \citeA{jacobsen2007io}. The main difference here is the different parameterization of the density, which was constant inside the torus with a strong gradient at the torus boundary in the case of \citeA{jacobsen2007io}. In our work, the reflections are less localized. 
 To investigate the energy fluxes that are available for particle acceleration and therefore for generation of aurora, we examine the energy flux through an analysis plane at $z=\pm60 R_{Io}$.

\subsubsection{Poynting Flux}

\begin{figure}
\begin{center}
\includegraphics[scale = 0.45]{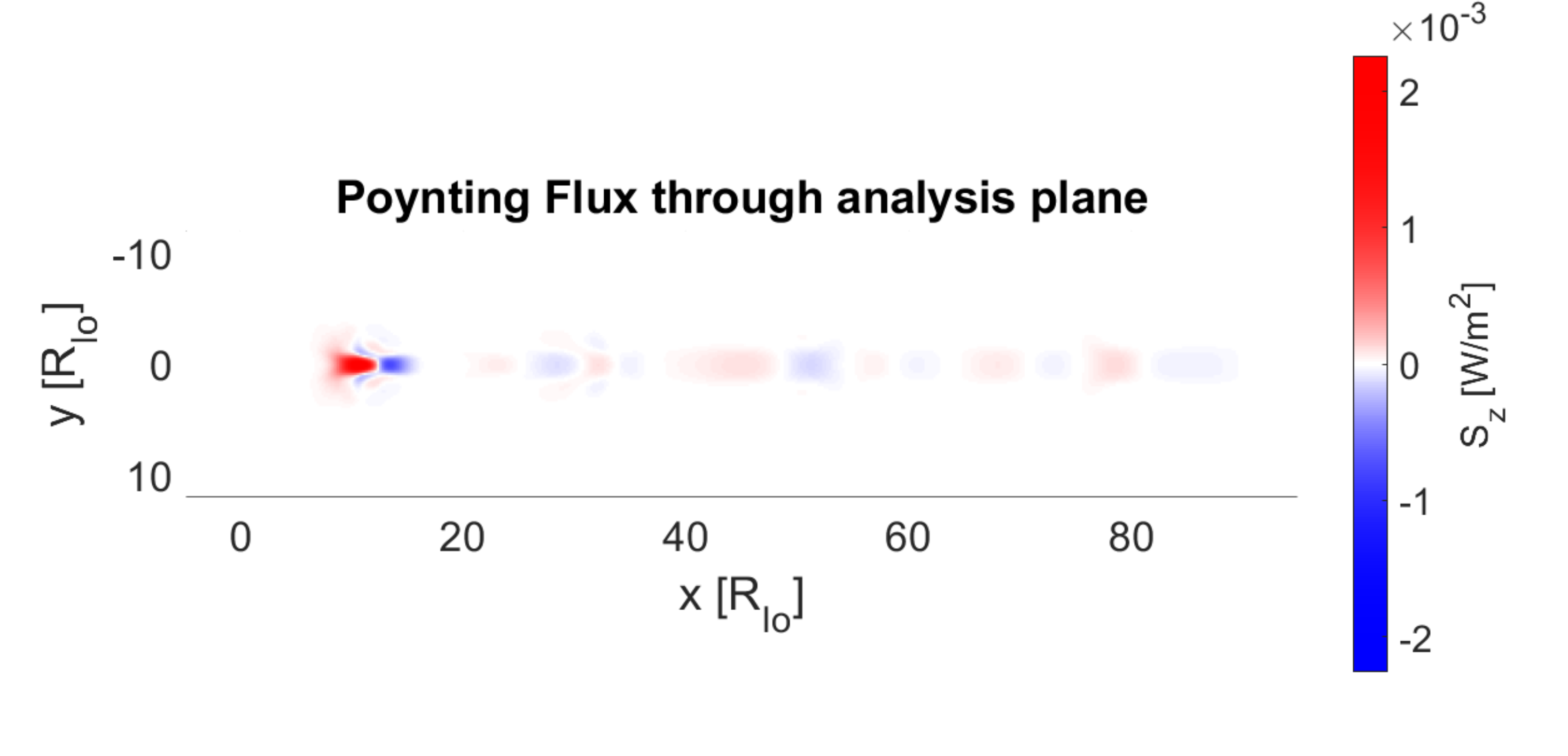}
\caption{The Poynting flux through the analysis plane near the northern Jovian ionosphere ($z = 60 R_{Io}$).  Positive (red) values shows flux towards Jupiter. The brightest spot at $x \approx 10-15 R_{Io}$ shows the incoming and reflected Poynting flux of the MAW. Multiple secondary Alfv\'en wings can be identified due to the reflections at the torus boundary and the Jovian southern and northern ionosphere.}
\label{fig:poyntingflux}
\end{center}
\end{figure}

Alfv\'en waves carry electromagnetic energy in the form of Poynting flux that can be partially 
 converted to particle acceleration near Jupiter \cite{hess2010power}. Since the MHD simulations do not include wave-particle interaction, we use the Poynting flux in the analysis plane as a proxy for morphology, position and maximum energy of the generated aurora. Even though the simulation is in the rest frame of Io, the Poynting flux is calculated in the rest frame of the plasma , i.e. in the rest frame of Jupiter, since we are interested in the energy flux generating imprints in Jupiter's atmosphere \cite{saur2013magnetic}. Here we are not interested in the direction of the Poynting flux, but only in the energy that is transported through the analysis plane,  since both the incoming as well as the reflected Poynting flux carry energy that can accelerate particles towards and away from Jupiter.
 \citeA{bonfond2009io} concluded that a mono-energetic electron distribution is not consistent with ultraviolet observations of the Io footprint tail. Broadband electron distributions and turbulent magnetic field fluctuations in the tail have been later confirmed by Juno observations
  \cite{szalay2018situ,szalay2020new, sulaiman2020wave, clark2020energetic} and suggest that wave-particle interaction of Alfv\'en waves is responsible for the particle acceleration (e.g. \citeA{saur2018wave,damiano2019kinetic}) in contrast to steady-state electric currents leading to a potential drop with mono-energetic, uni-directional acceleration.
 To more easily identify the reflection pattern and to interpret the results, we however, maintain the sign of the Poynting flux in the figures. Exemplarily, Figure \ref{fig:poyntingflux} shows the Poynting flux through the analysis plane of the reference model. The main Alfv\'en wing and its reflection from the Jovian ionosphere are well visible at $x \approx 10 R_{Io}$ and $x \approx 15 R_{Io}$, respectively. Further reflections due to the torus boundaries as well as the northern and southern Jovian ionospheres are visible. In the model the y-axis denotes the position alongside the connection line between Jupiter and Io, while negative values denote positions closer to Jupiter and positive values are further away from Jupiter. In the analysis plane however, the negative values can be interpreted as equatorward displacement and positive values as poleward displacement of the emission features in Jupiter's ionosphere from the corresponding central emissions.

\section{Interpretation of the Juno observations}

\begin{figure}
\begin{center}
\includegraphics[scale = 0.5]{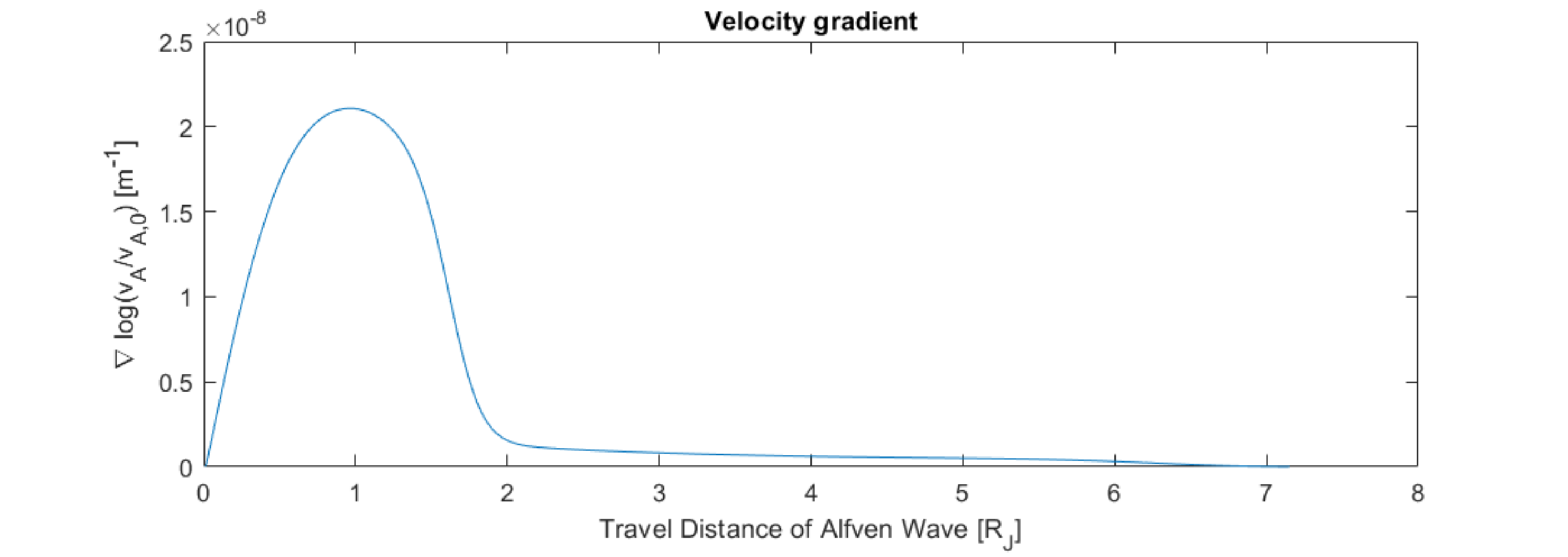}
\caption{Gradient of the Alfv\'en Wave velocity along the travel path starting in the center of the torus. This profile changes depending on the field line due to the position of Io inside the torus and the magnetic field strength in Io's vicinity. The maximum absolute gradient is at $\approx$~1~$R_J$, which is chosen as reflection point for the Alfv\'en wave at the torus boundary for the calculation of the location of the reflected waves in Figure \ref{fig:Mura}.}
\label{fig:VelocityGradient}
\end{center}
\end{figure}

 Figure \ref{fig:Mura} A shows observations by \cite{mura2018juno} taken with the JIRAM instrument on the Juno spacecraft. The observed IFPT contains a variety of substructures and multiple local brightness maxima, i.e. tail spots.  To evaluate whether the observed spots in the tail correspond to reflected Alfv\'en waves we conducted a study where we assigned these tail spots to different reflection patterns. For that purpose, we use the density model of \cite{dougherty2017survey}, the JRM09 magnetic field model \cite{connerney2018new} and the position of Io in the torus corresponding to the observations by \citeA{mura2018juno} in Figure \ref{fig:Mura} to calculate the travel times of Alfv\'en waves generated at Io on different magnetic field lines for a multitude of reflection patterns (Figure \ref{fig:Mura} C). The calculated corresponding footprint positions are shown in B. We omitted TEBs and the leading spot in this study. The position of Io for this study was calculated back from the location of emission of the Io footprint and is a few degree in longitude more downstream than Io's position at the time the image was taken. Therefore, each reflection pattern has a different relative position of Io inside the torus and therefore slightly different travel times towards the ionosphere and the torus boundaries. This reflection pattern is simplified in the sense that on the one hand, all reflections are linear and therefore no interaction between incident and reflected wave package is assumed (\citeA{jacobsen2007io}). Furthermore the reflections are assumed to be at distinct positions, one at the surface of Jupiter and on at the torus boundary. The torus boundary here is calculated as the largest gradient of the Alfv\'en velocity along the path as shown in Figure \ref{fig:VelocityGradient}.

As can be seen in Figure \ref{fig:Mura}, the position of the MAW spot and secondary spots calculated from this simplified wave pattern model map well to a range of secondary spots in the observations. Keeping the simplifications made and the dependence of the wave pattern on the underlying density model (e.g. \citeA{phipps2018distribution,dougherty2017survey}) in mind, the tail spots are consistent with reflections of Alfv\'en waves at torus boundary and ionosphere. 

\section{Origin of Alternating Alfv\'en Spot Street}

Now we discuss three possible mechanism which could break the observed symmetry in the Io footprint tail and could produce structures similar to the observed alternating Alfv\'en spot street. For every mechanism the Poynting flux through the analysis plane is compared to the Poynting flux through the same plane in the reference simulation.

\subsection{The Hall Effect} \label{sec:Hall}

To implement Hall conductivity in Io's ionosphere, we added the Hall term in the induction equation \ref{eq:induction}
\begin{equation}
\frac{\partial \mathbf{B}_{Hall}}{\partial t}  = -\frac{m}{\mu_0 e}\nabla \times \left(\left( \frac{\nabla \times \mathbf{B}}{\rho} \right) \times \mathbf{B} \right)
\end{equation}
 Single fluid MHD equations are equations for the mass density $\rho$ of a plasma and do not depend on the mass $m$ of an individual particle. However, the mass of the ions appear in the Hall term and in the calculation of the cross section \cite{saur1999three}.  We can therefore adjust the strength of the Hall effect by varying the particle mass in our simulation while keeping the mass density $\rho$ in all of the MHD equations unaffected and investigate the results for different ratios of Hall conductance to Pedersen conductance. Alternatively, we could change the neutral gas density of Io while maintaining the mass density, hence affecting the effective collision frequency. This approach however influences the interaction strength strongly, which renders the separation between effects that are related to the interaction strength and that to the Hall-effect difficult. It also leads to numerical problems for large neutral densities. Combining equation \ref{eq:pedersen_con} and \ref{eq:hall_con}, we can calculate the ratio $r$ of Hall conductance to Pedersen conductance as
\begin{equation}
    r = \frac{\sigma_H}{\sigma_P} = \frac{\nu}{\omega} = m \frac{\nu}{e B}
\end{equation}
For our reference model mass of $m = 24$~amu and the applied magnetic field strength and chosen neutral density, this results in a ratio of $r \approx 1/6$. 
The ratio reaches 1 at a particle mass of $m \approx 146$~amu. It is important to note that the different masses only act as weighting for the Hall effect while the plasma mass density stays unaffected. 
The Hall effect influences the electric current patterns in Io's gas cloud as the current is now not only along the electric field, but also  perpendicular to $\mathbf{E}_\bot$ and $\mathbf{B}$ and therefore roughly along the incoming plasma flow.  This results in a twist of the Alfv\'en wing and the symmetry between the Jovian and anti-Jovian side is broken \cite{saur1999three}. 

\begin{figure}
\begin{center}
\includegraphics[scale = 0.7]{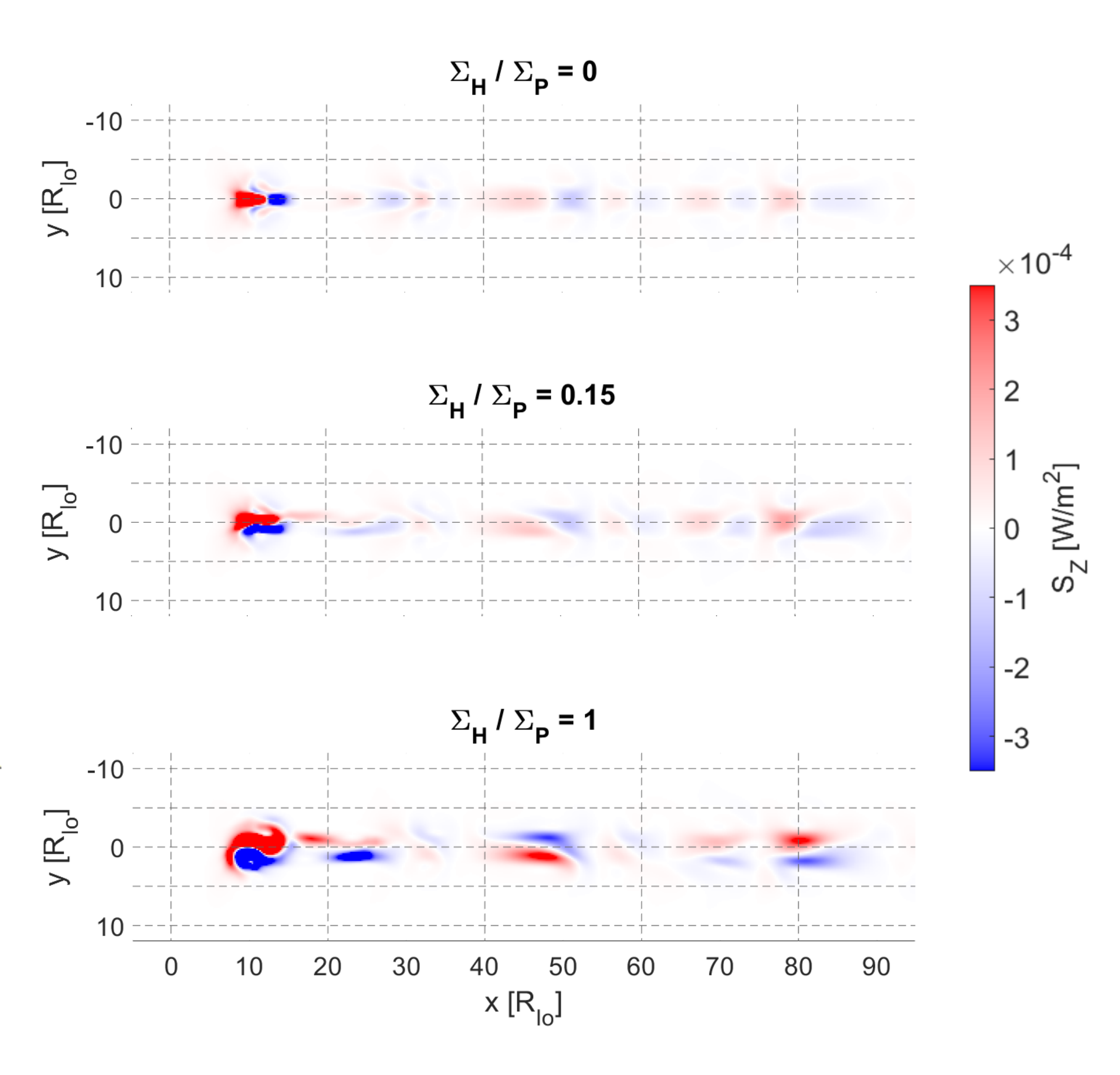}
\caption{Poynting flux of the MAW and the tail through the analysis plane for Hall to Pedersen conductance ratios of $r$~=~0,~0.15 and 1. The color bar has been adjusted to highlight features in the tail region. With higher Hall conductance, the asymmetry in the MAW and tail region increases. The maxima of the Poynting flux are laterally displaced and incoming and reflected Poynting fluxes are alternatingly displaced in positive (equatorward) and negative (poleward) y direction. Another effect of increased Hall conductance is the overall enhancement in intensity and spatial extend of the structures.}
\label{fig:Hall}
\end{center}
\end{figure}

To analyze the influence of the Hall effect in detail, we chose three different ratios $r$ of $0, 0.15$ and $1$ in our model. First, we will examine the influence of the Hall effect on the Poynting flux of the MAW through the analysis plane. Figure \ref{fig:Hall} shows the Poynting flux of the incoming (red) and reflected (blue) Alfv\'en wings through the analysis plane at $z = 55 R_{Io}$. As it can be seen, the symmetry along the path of the footprint, i.e. the y~=~0 plane,  breaks with increasing Hall effect strength. The morphology of the Poynting flux of the MAW at $x \approx 10 R_{Io}$ twists and the reflected and incoming Poynting flux can be seen at the same x position. This means that the maxima of incoming and reflected Poynting flux are laterally displaced. 

 The tail fluxes, i.e. the Poynting flux at $x > 20 R_{Io}$, are about one order of magnitude lower than the main fluxes. With increasing strength of the Hall effect, the asymmetry in the morphology of the tail Poynting flux grows. The maxima also move away from the y = 0 plane and therefore towards more distant latitudes from each other. This produces patterns which are similar in appearance to the observed alternating Alfv\'en spot street by \citeA{mura2018juno}. Also interesting to note is the positional displacement of incident and reflected Alfv\'en wing. The maxima of the reflected Alfv\'en wing and incoming Alfv\'en wing have no preferred displacement towards equator or towards the poles. If the particle acceleration due to incoming and reflected Poynting flux differ from each other, this could increase the alternating pattern of the emissions.

As a test for the importance of the Hall effect in breaking the symmetry, we can compare  its role on the footprint tails of the neighboring moons.  An important factor of the effect of the Hall conductance on the Poynting flux in the analysis plane are the multiple non-linear reflections which reinforce the asymmetry of the interaction. This means that the effect is weakened not only with lower Hall Conductance, but also with lower interaction strength in general. For Europa this would lead to a lesser development of the alternation and asymmetry in the Poynting flux morphology, but might still be observable. For Ganymede on the other hand, we would expect this effect to be negligible, since the Alfv\'en wing is generated by the interaction between the incoming plasma flow and Ganymede's magnetosphere and the ratio between Hall and Pedersen Conductance is very low \cite{kivelson2004magnetospheric,hartkorn2017induction}.   Since this would be in line with the new observations by \citeA{moirano2021morphology}, where no notable symmetry breaking was observed for Europa and Ganymede, we conclude that the Hall effect could play a major role in the creation of the observed alternating Alfv\'en spot street in Io's footprint tail.
%Asymmetries there might be due to the inner magnetic dipole and the external Jovian magnetic field not being aligned.

\subsection{Anti- and sub-Jovian Alfv\'en Wave Travel Time Difference} \label{sec:TravelTime}

\begin{figure}
\begin{center}
\includegraphics[scale = 0.4]{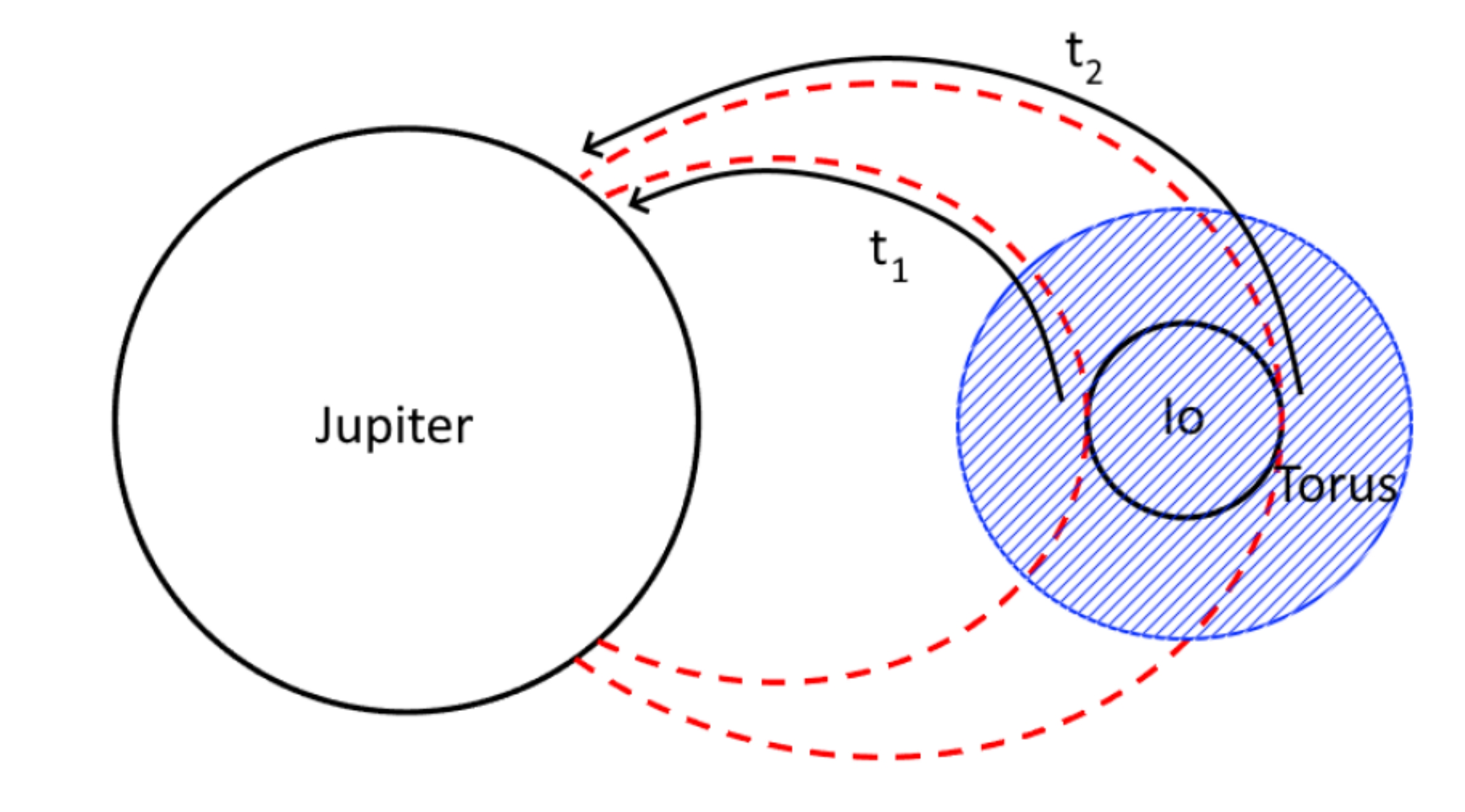}
\caption{Schematic of the travel time differences between Alfv\'en waves depending on their starting position. The travel path of the anti-Jovian Alfv\'en waves is longer than the travel path of the sub-Jovian Alfv\'en waves, resulting in a larger travel time $t_2 > t_1$. Since Io is not always centered in the torus, the difference in travel time varies.}
\label{fig:TravelTimeDifference}
\end{center}
\end{figure}

\begin{figure}
\begin{center}
\includegraphics[scale = 0.7]{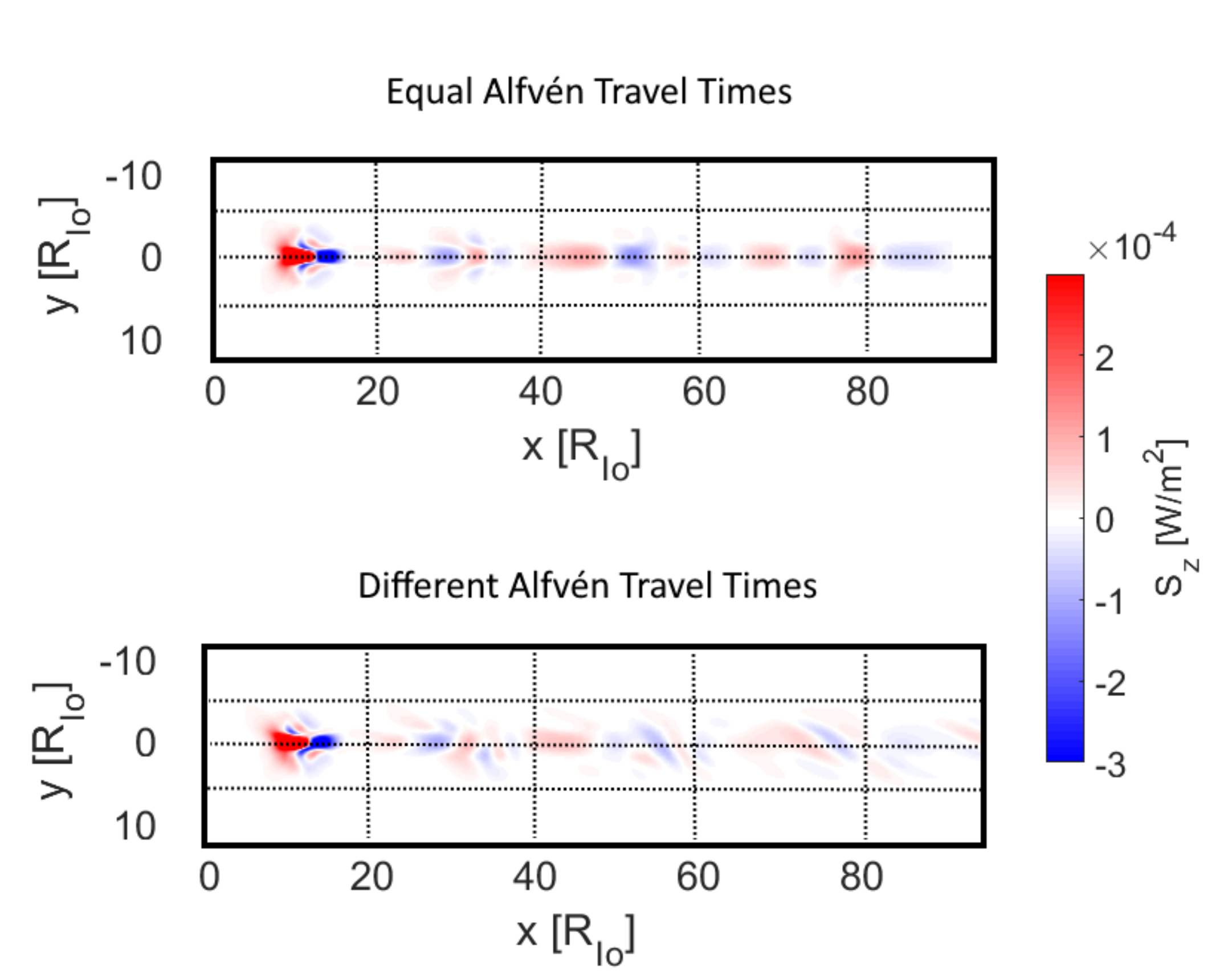}
\caption{Top: Poynting flux through the analysis plane of the reference model. Bottom: Poynting flux through the analysis plane of the model with travel time difference of 3.7~s. The effect is small at and near the MAW, but increases downstream. This results in asymmetry and displacement of the Poynting flux maxima pole- and equatorward.}
\label{fig:TravelTimes}
\end{center}
\end{figure}

The Alfv\'en waves that are generated at Io travel along their respective field lines towards Jupiter. Those generated at the Jupiter facing side of Io have a shorter travel path towards Jupiter than those generated at Io's anti-Jovian side (Figure  \ref{fig:TravelTimeDifference}). Depending on the density model (e.g. \citeA{bagenal1983alfven, dougherty2017survey, phipps2018distribution}) and position of Io inside the torus, we calculated this travel time difference to be between 1~s and 4~s These calculations are based on the JRM09 magnetic field model \cite{connerney2018new} and the density model by \citeA{dougherty2017survey} for different positions of Io in its orbit. Since the reference model is symmetric with respect to the y = 0 plane, no difference of the travel times from Io towards Jupiter occurs.  \\
To investigate the effects of the travel time difference that could influence the reflection pattern and morphology of the footprint, we implemented a density model that causes a 3.7~s delay between Jovian and anti-Jovian part of the Alfv\'en wing.  For the new density model $\hat{\rho}(y,z)$, we modify the density model given by equation \ref{eq:density} by adding a density gradient in y direction, which is set to $\lambda = 0.01 / R_{Io}$. 
\begin{equation}
    \hat{\rho}(y,z) = \rho(z) \cdot (1 + \lambda y)
\end{equation}
The comparison in Poynting flux through the analysis plane between a model with travel time difference and the reference model is shown in Figure \ref{fig:TravelTimes}. The 3.7~s travel time difference results in the plasma moving about $\delta x \approx 0.12 R_{Io}$ further downstream, before the Alfv\'en wave generated at the antijovian side of Io is propagated towards Jupiter and this side of the Alfv\'en wing intersects the analysis plane. This skews the MAW as seen in the lower plot. In the model, the travel time difference from Io to the torus boundary at $z = 25 R_{Io}$ is about 1.8~s while the travel time difference between torus boundary and Jovian ionosphere is 1.9~s. This results in different accumulated travel time differences in the IFP tail depending on where the reflections take place and how many occur. Even though the effect of the different travel times is small near the MAW, this effect can strongly contribute to the asymmetry of the Poynting flux in the tail region and create alternately displaced spots downstream of the MAW. Since the symmetry breaking and alternating spots are already visible close to the MAW in the observations by \citeA{mura2018juno}, the travel time difference  can likely not produce the observed structures by itself, but could be an amplifying effect. 

\begin{figure}
\begin{center}
\includegraphics[trim = 2cm 8cm 1.5cm 7cm, clip = true ,width = 1\textwidth]{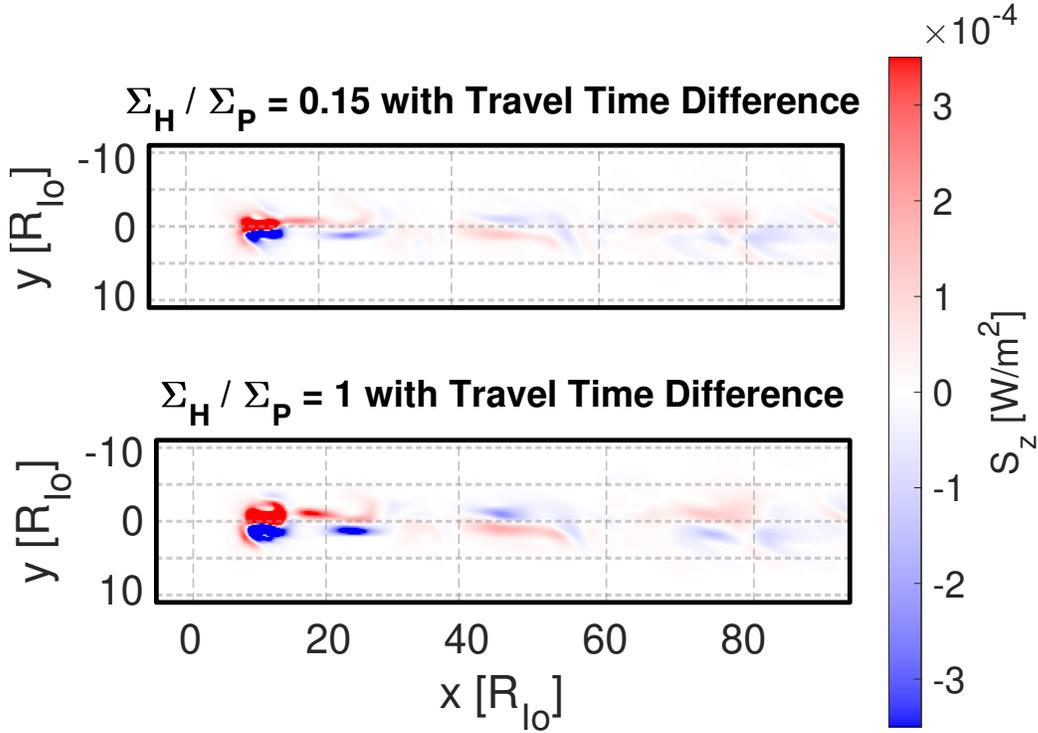}
\caption{Poynting flux through the analysis plane for a model with a travel time difference of 3.7~s and a Hall to Pedersen conductance ratio of $r = 0.15$ (top) and $r = 1$ (bottom). Near the MAW, the Poynting fluxes are comparable to those of the simulations without travel time difference, shown in Figure \ref{fig:Hall}. Further downstream they differ and the simulations with travel time difference show more substructure.}
\label{fig:HallAndTraveltime}
\end{center}
\end{figure}

To test this hypothesis, we carried out simulations with activated Hall effect and a travel time difference between Io's sub-jovian and anti-jovian side of 3.7~s. The results are shown in Figure \ref{fig:HallAndTraveltime}. We can now compare the Poynting fluxes through the analysis plane of these simulations with the ones without travel time difference, which are shown in the bottom two panels in Figure \ref{fig:Hall}. Generally, the travel time difference has a minor effect on the Poynting flux near the MAW, where the Hall effect already shows latitudinal displaced maxima. Further downstream however, its effect becomes more apparent. For an intermediate Hall conductance of $\Sigma_H = 0.15 \Sigma_P$, the Hall effect and travel time difference both contribute to the asymmetries in the Poynting flux and enhance the development of alternating maxima. Whereas the extrema in the Poynting flux in the simulation without travel time difference are still mainly at the same y position (latitude) further downstream, e.g. at x~=~40~$R_{Io}$, they feature a stronger latitudinal displacement in the simulations with travel time difference. The number of distinct maxima increase, albeit they appear to be weaker in value. In the case of a stronger Hall effect, where the Hall and Pedersen conductance are similar, the extrema in the Poynting flux already are heavily displaced in latitudinal direction and follow an alternating pattern. The effect of travel time difference in that case is only minor and does not significantly contribute any more to the development of the alternating Alfv\'en spot street.

The travel time difference depends on the position and size of the moon as well as the Alfv\'en velocity along the magnetic field lines. In Io's case, the travel time difference is especially high, because of the high density of the Io plasma torus, which increases the overall travel time. For larger M-shells, the travel time difference becomes smaller. For Europa, which is similar in size but farther outside and Ganymede which has a much larger interaction cross section, but is even further away from Jupiter, we do not expect the travel time difference to produce comparable patterns.  Since the alternating Alfv\'en spot street was observed in Io's footprint tail but not in those of Europa or Ganymede, the relevance of the travel time difference at Io compared to Europa and Ganymede is thus consistent with the observations of \citeA{moirano2021morphology}.
\subsection{Io's Asymmetric Atmosphere}\label{sec:Atmosphere}

\begin{figure}
\begin{center}
\includegraphics[scale = 0.6]{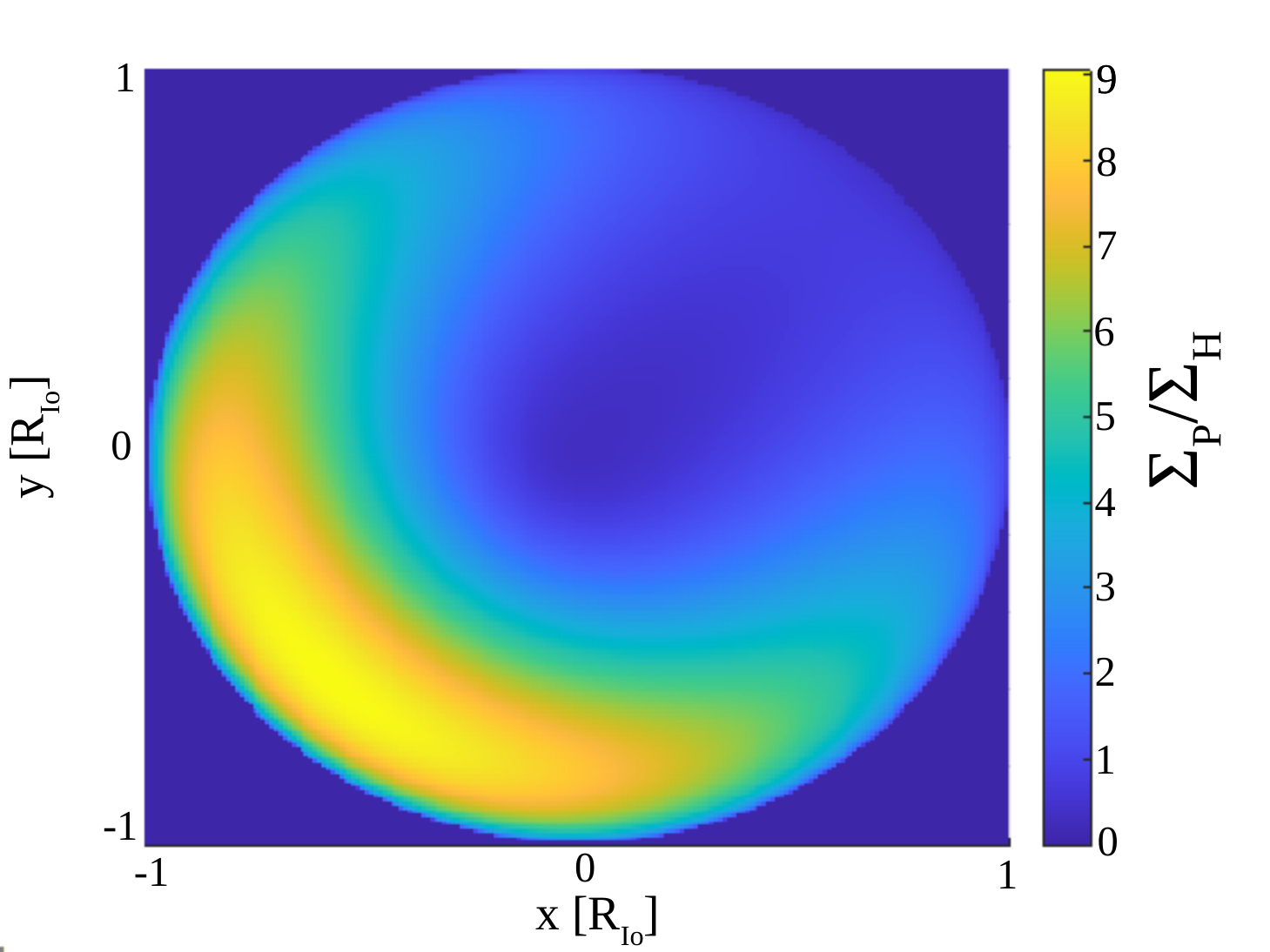}
\caption{Ratio of Pedersen conductance, integrated along the unperturbed magnetic field line to the Alfv\'en conductance at Io  for the implemented asymmetric atmospheric model. Since the atmospheric column density is lower at the poles while the integration length of the Pedersen conductivity is increased at the equator, the integrated polar Pedersen conductance is decreased, resulting in a lower ratio of about $\Sigma_P / \Sigma_A = 0.5$. Furthermore the used atmospheric model has a local maximum at the anti-Jovian ($y < 0$) trailing ($x < 0$) side, resulting in a high ratio of up to 9.}
\label{fig:AtmosphericModel}
\end{center}
\end{figure}

\begin{figure}
\begin{center}
\includegraphics[scale = 0.45]{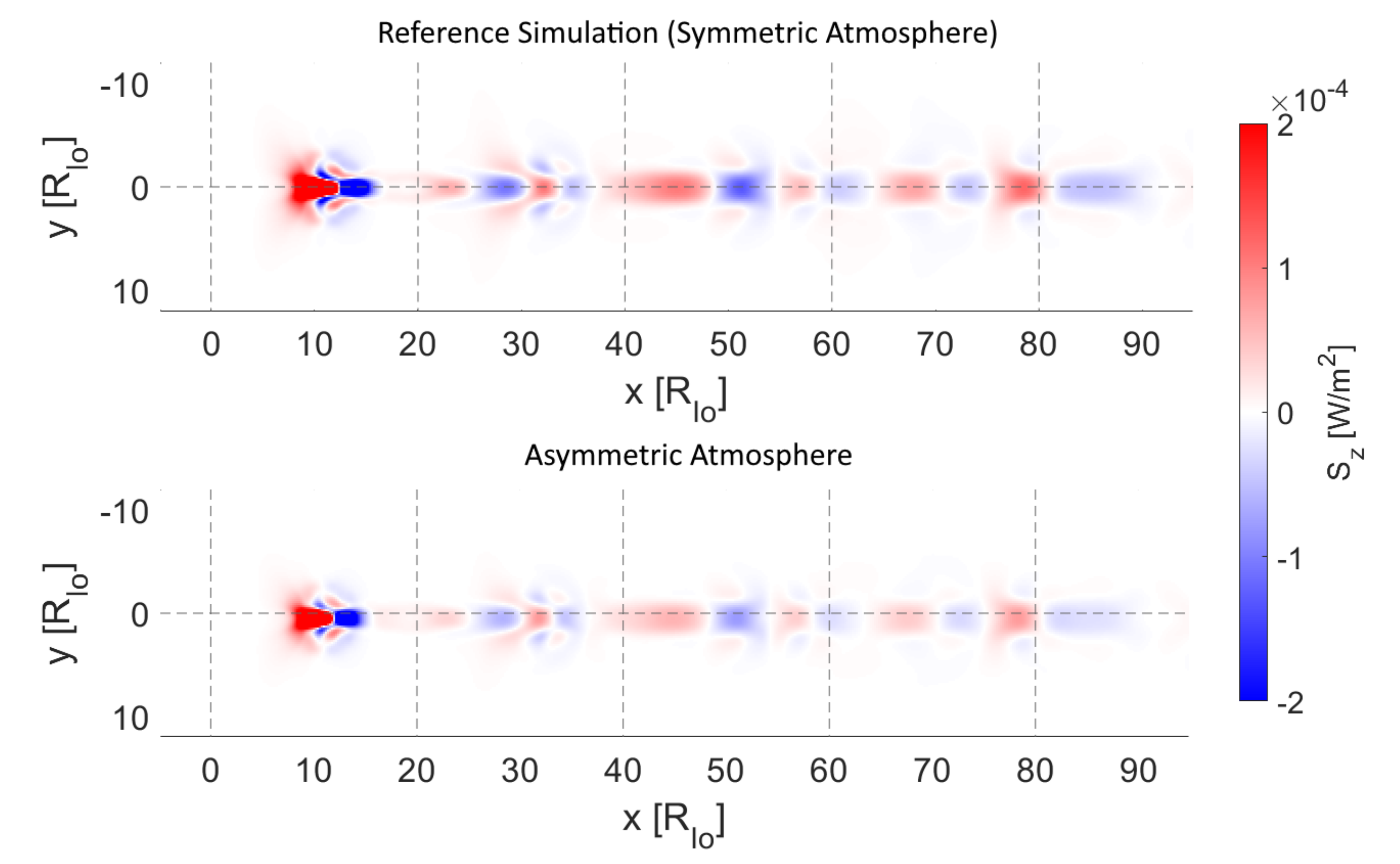}
\caption{Comparison of the Poynting flux through the analysis plane between the model with atmospheric asymmetry (top) and the reference model (bottom). Due to the shift in the maximum of the Pedersen conductance from (0,0) to approximately (-0.8, 0.8), the Poynting flux is shifted with respect to the symmetry axis of the reference model at y=0 (dashed line). Asymmetries with respect to a shifted symmetry axis at approximately y = 0.8 in the Poynting flux are negligible.}
\label{fig:AtmosphericAsymmetry}
\end{center}
\end{figure}

Io's atmosphere is highly asymmetric between day and night side as well as leading and trailing and between sub-Jovian and anti-Jovian side \cite{lellouch2007io}. Further asymmetries include a much higher atmospheric column density at the equator than at the poles. Since the atmosphere is not only created by sublimation but also by volcanic out-gassing generating additional local density, the atmosphere is likely patchy with local maxima. A patchy and asymmetric atmosphere possibly leads to a strong variation of Pedersen conductance and therefore in interaction strength. Due to the non-linear reflections that occur at these high interaction strengths \cite{jacobsen2007io}, this might result in an asymmetric reflection pattern and therefore breaking the symmetry between sub- and anti-Jovian side.  To investigate, whether the atmosphere could break the symmetry, we used a asymmetric atmosphere with a wide range of field line integrated Pedersen conductances $\Sigma_P(x,y)$.  The atmospheric model has an increased column density at the equator and a local increase at the trailing, anti-Jovian side.   To get an estimate, how the inhomogeneous atmosphere will affect the flow inside of the Alfv\'en wing, we look at the ratio between the Pedersen conductivity integrated along magnetic field lines $\Sigma_P(x,y)$ for different magnetic field lines at the position $(x,y)$ to the Alfv\'en conductance $\Sigma_A$. Similar to equation \ref{eq:interaction_strength}, higher values represent a stronger local perturbation of the plasma flow and magnetic field.  The calculated distribution of the interaction strength is shown in Figure \ref{fig:AtmosphericModel}.
The Poynting flux through the analysis plane caused by this atmospheric model is shown in Figure \ref{fig:AtmosphericAsymmetry}.  Most notable is the shift of location in y direction of the complete Poynting flux pattern, which is due to the shift in local maxima of the interaction strength as seen in Figure \ref{fig:AtmosphericModel}.  Small scale structures of the atmosphere are not visible in the Poynting flux. The effect might be further diminished by  numerical   dissipation along the Alfv\'en wave travel path. However, even without numerical dissipation, the symmetry breaking effect of the atmosphere seems to be minor and has no noticeable effect for symmetry breaking downstream of the main spot in the tail.

\section{Summary \& Conclusion}

Juno observations by \citeA{mura2018juno} and \citeA{moirano2021morphology} show the Io footprint tail with multiple tail spots that show a pattern of alternatingly displaced secondary spots. Calculating travel times of the Alfv\'en waves generated by Io for different reflections patterns show that the secondary spots could be generated by Alfv\'en waves that are reflected at the torus boundary and Jupiter's ionosphere. On this basis, we presented a study of three mechanisms of the Io Jupiter interaction using single fluid Hall-MHD simulation to explain the observed structures. The model includes Io as a neutral gas cloud generating Alfv\'en waves that travel along magnetic field lines towards Jupiter, where we used the Poynting flux as a proxy for the morphology, position and strength of the auroral emissions in the Io footprint and its tail. For the reflection at the torus boundary and Jupiter's ionosphere a density gradient along the field lines is implemented while matching the estimated total travel times of Alfv\'en wave along the field lines. We conclude that two of the three investigated effects produce or enhance symmetry breaking with the Hall-Effect being the most promising for generating the observed alternating Alfv\'en spot street in the Io footprint tail. \\
With a parameter study for different Hall to Pedersen conductance ratios, we show that the Hall effect strongly modifies the structure of the main footprint and the location and structure of the secondary spots. Especially with a high Hall-Conductance (Hall to Pedersen conductance ratio = 1) as expected by \citeA{saur1999three} and \citeA{kivelson2004magnetospheric}, the symmetry along the wake breaks down and local Poynting flux maxima  are displaced pole- and equatorwards. Therefore, the Hall effect is a strong candidate to produce alternating footprints. Furthermore, the Hall effect is weaker at Europa and Ganymede, where no similar structures have been observed \cite{moirano2021morphology}. To test this hypothesis further, additional high resolution observations of the footprints of Europa and Ganymede are necessary. A similar pattern in the tail of the other Galilean moons could likely be not explained by the Hall-effect alone. \\
We further investigated the effect of the travel time difference of the Alfv\'en waves starting from the sub-Jovian and anti-Jovian side of Io on the IFPT morphology. Although the effect produces strong symmetry breaking further down the footprint tail and grows with distance, the Poynting flux is only weakly disrupted near the MAW. In the observations however, the symmetry breaking is already visible close to the main footprint emissions. Therefore, we argue that the travel time difference can not produce the observed pattern by itself, but could be a contributing effect further downstream. Since the travel time difference is smaller at Europa and Ganymede, we do not assume it to be a notable effect at the other Galilean moons. \\ As the third effect we investigated the influence of an asymmetric atmosphere on the morphology of the footprint and its tail. We find no evidence that this effect contributes notably to symmetry breaking and rule out inhomogeneities in the atmosphere as the reason  for the observed pattern.  \\ 
We conclude that the most promising effect to create the observed alternating Alfv\'en spot street in the Io footprint tail is the Hall effect. This could be further tested by new observations of the footprints of the Galilean moons. If the effect is not visible in the footprint tails of Ganymede and Europa, this would be consistent with  the Hall effect as the primary reason behind the alternating Alfv\'en spot street.

\section{Open Research}

All data in this study are created using the PLUTO MHD code version 4.4, available at \url{http://plutocode.ph.unito.it/download.html}. All changes to the code are specified in the manuscript in Section \ref{sec:MHDModelEquations}. The used boundary and initial conditions as well as the grid are described in Section \ref{sec:Model} and all values for the background model and paramterization of Io's neutral gas cloud are given in Table \ref{tab:properties}.

\acknowledgments
 We thank the Regional Computing Center of the University of Cologne (RRZK) for providing computing time on the DFG-funded (Funding number: INST 216/512/1FUGG) High Performance Computing (HPC) system CHEOPS as well as support. \\
 This project has received funding from the European Research Council (ERC) under the European Union’s Horizon 2020 research and innovation programme (grant agreement No. 884711).

\

\bibliography{agusample}

\end{document}